\documentclass[aps,prb,twocolumn,superscriptaddress,a4paper]{revtex4}
\usepackage{graphicx}
\usepackage{amsmath}
\begin{document}
\title{\bf Spin-Transfer Torque in Helical Spin-Density Waves}
\author{O. Wessely}
\affiliation{Department of Mathematics, Imperial College, London SW7 2BZ, United Kingdom}
\affiliation{Department of Mathematics, City University, London EC1V 0HB, United Kingdom}
\affiliation{Department of Physics, Uppsala University,
Box 530, SE-75121, Uppsala, Sweden}
\author{B. Skubic}
\affiliation{Department of Physics, Uppsala University,
Box 530, SE-75121, Uppsala, Sweden}
\author{L. Nordstr\"om}
\affiliation{Department of Physics, Uppsala University,
Box 530, SE-75121, Uppsala, Sweden}
\date{\today }

\begin{abstract}
The current driven magnetisation dynamics of a helical spin-density wave is investigated. Expressions for calculating the spin-transfer torque of real systems from first principles density functional theory are presented. These expressions are used for calculating the spin-transfer torque for the spin spirals of Er and fcc Fe at two different lattice volumes. It is shown that the calculated torque induces a rigid rotation of the order parameter with respect to the spin spiral axis. The torque is found to depend on the wave vector of the spin spiral and the spin-polarisation of the Fermi surface states. The resulting dynamics of the spin spiral is also discussed.
%The presence of a current induced torque in a one-dimensional model of a helical spin density wave is shown. 
\end{abstract}
\maketitle

\section{Introduction}
\noindent The spin-transfer torque (STT) provides a means of manipulating the magnetisation of a material using a current. The effect was proposed from theoretical considerations by Slonczewski\cite{Slon} and Berger\cite{Berg} and has attracted a lot of attention due to its potential use in applications where a magnetic state is altered by a current as opposed to traditional techniques involving magnetic fields.

Most theoretical and experimental work on the STT concerns layered magnetic structures where the STT appears as a consequence of angular momentum conservation as a current traverses an interface between two regions of non-parallel magnetisation. Recently\cite{Wessely2006} we have shown that a STT also occurs in systems with a helical spin-density wave (SDW). This was shown by a first-principles calculation of the STT for a bulk rare-earth system (Er) in its low temperature helical spin spiral (SS) structure. The effect of the current induced STT in Er is a rigid rotation of the SS order parameter. This phenomenon is a bulk effect in contrast to STT in 
layered systems where the STT mainly occurs close to the interfaces between the layers.
%On an atomic level Er can be seen as a layered structure where the magnetic order of each monolayer is slightly rotated forming the spin spiral structure. 

In this Article we present a more general discussion on the current induced torque in a SS. First the effect is illustrated using a one-dimensional model of a SS. We in the subsequent section present first-principles calculations of the STT for three real systems, Er and fcc Fe or $\gamma$-Fe at two different lattice volumes. Previous first-principles calculations have shown that both Er and $\gamma$-Fe have helical 
SDWs.\cite{Mryasov1992,Bettan} In many other respects these two systems are very different. Erbium is one of several rare-earth elements that exhibit a helical SDW and where the ordering is driven by a nesting between parallel sheets of the Fermi surface (FS). Iron belongs to the 3d-transition metals where SDWs 
are less frequent. The fcc phase of Fe, which is the phase where a SS order has been found both in experiments and theory, has only been stabilised at very special conditions, and it is believed that the helical SDW
is the combined result of several parts of the FS. For the one-dimensional model, as well as the considered real systems, we find that an applied current induces a STT generating a rigid rotation of the spiral order. With this theoretical prediction, new types of potential applications of the STT can be imagined, such as current driven oscillators with tunable frequencies. The current driven spin dynamics for such a device is investigated in 
section (\ref{sec:Dyn})
\section{Theory}
\begin{figure}
\includegraphics[scale=0.5]{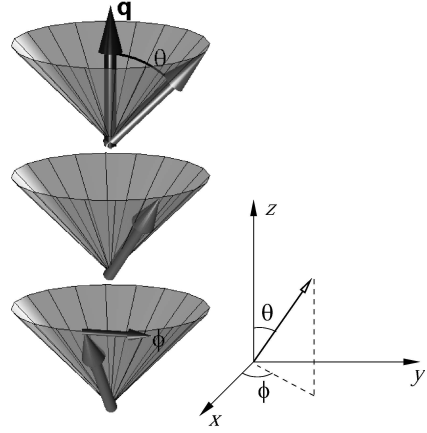}%
\caption{\label{Sp} Local magnetisation direction of a helical spin density
wave. The spin density wave is characterised by the SS wave vector ${\bf q}$, the cone angle $\theta$, and the rotation angle $\phi={\bf q}\cdot{\bf r}$.}
\label{fig:coord}
\end{figure}
\noindent For a system with a helical SDW, the direction of the local magnetisation rotates around the SS quantisation axis as one moves in the direction of the SS wave vector. Without loss of generality, we here consider the rotation axis to be parallel to the SS wave vector, except when explicitly stated, and will refer to it as the SS axis. Although we here consider helical SDWs, the results are valid for cycloidal SDWs as well. A SS is characterised by its SS wave vector ${\bf q}$ and its cone angle $\theta$, which is the angle between the SS axis and the local magnetisation, see Fig.~\ref{fig:coord}. In this section we first consider a one-dimensional model of a SS, where the spin currents and STT are calculated analytically. We then discuss how accurate first-principles calculations of the STT can be performed. 

\subsection{The STT in a 1D model of a spin spiral}
\label{sec:1D}
\noindent As a simple one-dimensional single band model of a SS system, we consider independent particles in a spin-dependent potential. The spin-dependent potential is chosen such that the ground state of the system has a 
helical magnetisation ${\bf M}=M{\hat{\bf m}}$ where
%can be constructed which generates a solution with a helical magnetization, 
\begin{equation}
{\hat {\bf m}}=(\textrm{sin}~\theta~\textrm{cos}~qz,~\textrm{sin}~\theta~\textrm{sin}~qz,~\textrm{cos}~\theta).
\label{eq:helicalm}
\end{equation}
Here, the SS wave vector ${\bf q}=q{\hat{\bf z}}$ and $\theta$ is the cone angle.
%of constant magnitude and with wave vector ${\bf q}=q {\hat {\bf z}}$.
The single particle Schr{\"o}dinger or Kohn Sham equation for such a model system takes the form
 %The Shr{\"o}dinger equation for an electron moving 
%in a one-dimensional system with a helical spin density wave is given by
	\begin{displaymath}
\begin{array}{cr}
\Big\{-{\hbar^2\over 2 m_e}{\partial^2\over\partial z^2} & +
{U \over 2}\big[\textrm{sin}~\theta~\left(\textrm{cos}~qz~{\sigma}_x+\textrm{sin}~qz~{\sigma}_y\right)\\
& +\textrm{cos}~\theta~{\sigma}_z\big] -\epsilon \Big\}\psi=0,
\end{array}
	\end{displaymath}
where $U$ is the exchange splitting, $m_e$ is the electron mass, $\hbar$ is the Planck constant $\sigma_x,\;
\sigma_y,\;\sigma_z$ are the Pauli matrices and $\epsilon$ is the one particle energy. This equation can be solved analytically\cite{OH} and the eigenfunctions are the so called generalised Bloch states,
	\begin{equation}
	  \psi_k=\left( \begin{array}{c} \psi_{nk}(z,\uparrow) \\
	    \psi_{nk}(z,\downarrow)\end{array} \right)=
	  e^{ikz}\left( \begin{array}{c} \textrm{cos}~(\theta_k/2)~
	    e^{-iqz/2}\\
	    \textrm{sin}~(\theta_k/2)~e^{iqz/2}\end{array} \right).
          \label{eq:blochstate}
	\end{equation}
For the above wave function the direction of the spin ${\hat{\bf s}}$ rotates around the SS axis, with rotation angle $\phi=qz$, at an angle $\theta_k$ with respect to the SS axis, 
\begin{equation}
{\hat{\bf s}}(s)=(\textrm{sin}~\theta_k~\textrm{cos}~qz,~\textrm{sin}~\theta_k~\textrm{sin}~qz,~\textrm{cos}~\theta_k).
\label{eq:spin}
\end{equation}
Note that the single electron cone angle $\theta_k$ and the cone angle $\theta$ of the SS, see Fig.~\ref{fig:coord}, in general are different.
%	\begin{center}
%	\includegraphics[width=0.8\textwidth]{Olafig/spiralband.eps}\\[8mm]
%	\end{center}
%	\vspace{-15mm}
In one dimension, the charge current $j$, induced by applying an electric field, is a scalar given by 
	\begin{equation}
	  j(z)=-\delta f{e \hbar\over m_e}Re\Big(\psi_{k_F}^*
	  i{\partial \over \partial
	    z}\psi_{k_F}-\psi_{-k_F}^*i{\partial \over \partial
	    z}\psi_{-k_F}\Big),
          \label{eq:current}
 	\end{equation}
where $\delta f$ is the change in occupation at the Fermi level due to the electric field, $k_F$ is the Fermi wave vector and $e$ is the elementary charge. In a similar way, the spin current ${\bf Q}$ due to a weak electric field is a vector given by  
 	\begin{equation}
	    {\bf Q}(z)=-\delta f{\hbar^2 \over 2 m_e} Re\Big(\psi_{k_F}^*i
	    {\partial \over \partial z}{\mathbf \sigma}\psi_{k_F}-
	    \psi_{-k_F}^*i{\partial \over \partial z}{\mathbf
	      \sigma}\psi_{-k_F}\Big).
          \label{eq:spincurrent}
	\end{equation}
From the spin current can the rate of change of angular momentum within an infinitesimal region of the system
be calculated.
%we may calculate the spin-flux or change of angular momentum, within an infinitesimal region of the system, 
The rate of change or torque on the region
between $z_0$ and $z_0+dz$ is given by the spin flux into the region
% spin flux $\partial{\bf J}/ \partial t$ into an infinitesimal region is given by
	\begin{equation}
	  {\partial {\bf J}\over \partial t}(z_0)=Q(z_0)-Q(z_0+dz)=-{\partial {\bf Q}\over
	    \partial z}\Big|_{z_0}dz.
	\end{equation}
%Since the model SS is uniform in space, the spin flux into $dz$ is independent of $z_0$. 
%Inte helt korrekt
Both the spin-flux $\partial {\bf J}/\partial t$ and the current $j$ depend on the change in occupation of
the states at the Fermi level caused by the applied electric field. These two quantities are linked by the expression,
	\begin{equation}
	  {\partial {\bf J}\over \partial t}= \textsf{C} j,
	\end{equation}
which defines the torque current tensor $\textsf{C}$.
The torque current tensor is in three dimensions a matrix, but becomes a
vector in a one-dimensional system, where also the current vector is reduced to a scalar. 
%the torque current tensor, $\textsf{C}$, becomes a vector. 
For our model system, the torque current tensor can be calculated by inserting the Fermi Bloch states, $\psi_{k_F}$ and $\psi_{-k_F}$ from Eq.~(\ref{eq:blochstate}) into Eq.~(\ref{eq:current})-(\ref{eq:spincurrent}), resulting in, 
% and for the SS system using the SS eigenfunctions and the expression for the angular momentum change and current in an infinitesimal region in the system we get 
	\begin{equation}
	  \textsf{C}={\partial {\bf Q}\over \partial z}\Big|_{z_0}dz/j(z_0)=  
	  a(\theta,k_F)\left( \begin{array}{c} -q~\textrm{sin}~(qz_0)~\\
	    q~\textrm{cos}~(qz_0)~\\
	    0\end{array} \right)dz,
 \label{eq:torquetensor}
	\end{equation}
where the first factor, $a(\theta,k_F)$, depends on the spiral cone angle $\theta$ and the Fermi wave vector $k_F$. For planar spin spirals, where $\theta=\pi/2$, the factor reduces to a simple function of the single electron cone angle $\theta_k$ of the states at the Fermi level given by 
	\begin{equation}
	  a(\pi/2,k_F)={\hbar \over 2e}~{\textrm{sin}~\theta_{k_F}\over
	    1-(q/2k)\textrm{cos}~\theta_{k_F}}. 
	  \label{eq:afactor}
	\end{equation}
%This is the spin torque tensor for any infinitesimal region in the system. For a planar spin spiral the function $a(\theta)$ is given by the spin polarization of the electrons at the Fermi surface 
%	\begin{equation}
%	  \textsf{C}\equiv {\partial {\bf J}_n\over \partial t}/j(z)=  
%	  \left( \begin{array}{c} -{\hbar \over e}~\textrm{sin}~\theta_{k_F}~\textrm{sin}~qz~q\\
%	    {\hbar \over e}~\textrm{sin}~\theta_{k_F}~\textrm{cos~qz~q\\
%	    0\end{array} \right)
%	    \nonumber
%	\end{equation}
%	 Ref [1] A. W. Overhauser Phys. Rev. B{\bf 128} 1437 (1962)
%\section{Dynamics}
A comparison between Eq.~(\ref{eq:helicalm}) and Eq.~(\ref{eq:torquetensor}) shows that 
%From the structure of the torque current tensor, $\textsf{C}$, we conclude that 
the current induced torque, in our one-dimensional model, is perpendicular to the magnetisation direction and lies in the plane of rotation of the SS, thus causing the SS to rotate with respect to the SS axis. 
%make a rigid rotation. 
It can also be seen from Eq.~(\ref{eq:torquetensor}) that the magnitude of the STT
is governed by the length of the SS wave vector $q$ and the factor $a(\theta,k_F)$.
%where  $a(\theta,k_F)$ for a planar spiral is given by the 
%spin polarization of the electrons at the fermi level.

\subsection{The STT for a real system}
\noindent The STT for a real system can be calculated by generalising the 
expressions for the charge and spin currents in one dimension to three 
dimensions. In three dimensions can the currents induced by 
%a relatively weak 
an external electric 
field, to linear order in the field,  be calculated from a FS integral. 
For a multiband system must the contribution from all bands crossing the Fermi level be taken into account. 
In the previous section was a model system and the STT on an infinitesimal region 
at an arbitrary point of the system considered. For a real material, where 
most of the angular momentum is localised on the atoms is the time evolution of the atomic
moment governed by the STT on the corresponding atom. 
The torque on an atom is given by the spin-flux 
into a sphere surrounding the atom. For a single electron state with band index $n$ and wave vector ${\bf k}$,
the flux into a sphere of radius $R$ is given by,
 	 \begin{equation}
	   -\int_0^\pi \int_0^{2\pi} {\bf Q}_{n{\bf k}}\cdot {\hat{\bf r}}
R~\textrm{sin}~\theta~d \phi d\theta ={\partial {\bf J}_{n{\bf k}} \over \partial t}.
\label{eq:flux}
	 \end{equation}
In three dimensions is the one electron spin current tensor%, ${\bf Q}_{n{\bf k}}$, is a tensor given by 
	 \begin{equation}
	   {\bf Q}_{n{\bf k}}({\bf r})=\mathrm{Re} \left\{\psi^\dagger_
	   {n{\bf k}}({\bf r})\,
	   {\bf s} \otimes{\bf v}\psi_{n{\bf k}}({\bf r})\right\},
	 \end{equation}
where ${\bf s}=(\hbar/2){\bf \sigma}$, and ${\bf v} =(-i\hbar/m_e)\nabla$.
The total spin current is the sum of the spin currents from all occupied states.
A external electric field ${\bf E}$ induces a non equilibrium spin current by changing the occupation of the states $\delta f$ at the FS. The change in occupation number $\delta f$ can be calculated using the relaxation time approximation and semiclassical Boltzmann theory,\cite{xiao}
\begin{equation}
\delta f({\bf k})=-\delta(\epsilon_{n{\bf k}}-\epsilon_F){\tau e \over \hbar} \nabla_{\bf k}\epsilon_{n{\bf k}}\cdot{\bf E},
\end{equation}
where $\tau$ is the electron relaxation time and $\epsilon_{n{\bf k}}$ is the band energy. The total 
non equilibrium torque on an atom can be calculated from a FS integral,
\begin{equation}
{\partial {\bf J}\over \partial t}=
{-V_C\tau e\over (2\pi)^3\hbar}\sum_{n}\int_{FS}
{\partial {\bf J}_{n{\bf k}}\over \partial t}
\left(\nabla_{\bf k}\epsilon_{n{\bf k}}
\cdot{\bf E}\right){dS_{n \bf k}\over|\nabla_{\bf k}\epsilon_{n{\bf k}}|} \, ,
\label{eq:A1}
\end{equation}
where $V_C$ is the volume of the system and a
summation is performed over all bands crossing the Fermi level. The above equation 
defines a linear relation between the torque and the external field,  
\begin{equation}
{\partial {\bf J}\over \partial t}=\tau \sum_{n} \textsf{A}_n{\bf E}.
%{\overline {\overline A}}_{n}{\bf E}.
\label{eq:A}
\end{equation}
A similar 
relation can be obtained for the charge current density and the electric field,
\begin{eqnarray}
{\bf j}&&=\tau \sum_n 
\textsf{B}
%{\overline {\overline B}}
_n{\bf E}\label{LRI}
\label{eq:B}\\
&&={1\over(2\pi)^3}{\tau e^2\over \hbar^2}\sum_{n}\int_{FS}
\nabla_{\bf k}\epsilon_{n{\bf k}}
\left(\nabla_{\bf k}\epsilon_{n{\bf k}}
\cdot{\bf E}\right){dS_{n \bf k}\over|\nabla_{\bf k}\epsilon_{n{\bf k}}|},
\nonumber
\end{eqnarray}
%
%Lars lite omformuleringar
where $\textsf{B}_n$ is the conductivity tensor for band $n$. By combining 
Eqs.~(\ref{eq:A})-(\ref{eq:B}) is a linear relation between the torque
and the current density obtained, where the unknown electron relaxation time $\tau$ 
has been cancelled,
\begin{equation}
{\partial {\bf J}\over \partial t}= (\sum_n\textsf{A}
%{\overline {\overline A}}
_n)(\sum_m \textsf{B}
%{\overline {\overline B}}
_m)^{-1}{\bf j}=
\textsf{C}
%{\overline {\overline C}}
{\bf j}.
\label{LRJI}
\end{equation}
The above equation defines the spin current tensor $\textsf{C}$ in 3 dimensions.
\subsection{The STT from the augmented plane wave (APW) method}
\noindent The wave functions used in the previous section to calculate the torque current 
tensor $\textsf{C}$ can be obtained from first principles electronic structure calculations
using spin density functional theory. 
Several electronic structure methods use basis sets where space is divided into muffin-tin 
spheres surrounding the atoms
and an interstitial region.
This division is introduced in order to simplify the calculation of the
% to account for the 
intra and interatomic behaviour of the wave functions. When calculating the STT, a natural choice is to use the atomic augmentation (muffin-tin) sphere as the surface for evaluating the STT on the atoms. We now illustrate more in detail how the calculation may be carried out within the augmented plane wave method (APW). The APW expansion can be written as a sum of plane waves
%\begin{eqnarray}
%\psi_{n{\bf k}}({\bf r})=\sum_{\bf G}
%\Big (\alpha a_{n{\bf k},\bf G}e^{i({\bf G}+{\bf k}-{\bf q}/2){\bf r}} \nonumber
%\\
%+\beta b_{n{\bf k},\bf G}e^{i({\bf G}+{\bf k}+{\bf q}/2){\bf r}}\Big )
%\end{eqnarray}
\begin{equation}
\psi_{n{\bf k}}({\bf r})=\sum_{\bf G}\left( \begin{array}{c} a_{n{\bf k},\bf G}e^{i({\bf G}+{\bf k}-{\bf q}/2){\bf r}}\\
b_{n{\bf k},\bf G}e^{i({\bf G}+{\bf k}+{\bf q}/2){\bf r}}\end{array} \right),
\end{equation}
where $\bf G$ are the reciprocal lattice vectors. The plane wave coefficients $a_{n{\bf k}}$ and $b_{n{\bf k}}$ are obtained from the first principles calculation. The spin-flux into a sphere with radius $R$ centred at an atom at site ${\bf r}_m$ is for plane waves given by the expression
\begin{widetext}
\begin{eqnarray}
\int_0^\pi \int_0^{2\pi} {\bf Q}_{n{\bf k}}\cdot {\hat{\bf r}}
R~\textrm{sin}~\theta~d \phi d\theta=&&{\hbar R^2\over m_e}\mathrm{Re}\,
\sum_{{\bf G},{\bf G'}}-i4\pi\Big[ \nonumber\\
&&\alpha^\dagger{\bf s}\alpha \; a^*_{n{\bf k},\bf G}
 a_{n{\bf k},\bf G'} e^{-i({\bf G}-{\bf G'}){\bf r}_m}j_1(|{\bf G}-{\bf G'}|R)\,
 ({\bf G'}+{\bf k}-{\bf q}/2){\cdot} ({\widehat{{\bf G}-{\bf G'}}})\nonumber\\
+&& \alpha^\dagger{\bf s}\beta\;a^*_{n{\bf k},\bf G}
 b_{n{\bf k},\bf G'} e^{-i({\bf G}-{\bf G'}-{\bf q}){\bf r}_m}j_1(|{\bf G}-{\bf G'}-{\bf q}|R)\,
({\bf G'}+{\bf k}+{\bf q}/2){\bf \cdot} ({\widehat{{\bf G}-{\bf G'}-{\bf q}}})\nonumber\\
+&&\beta^\dagger{\bf s}\alpha\;b^*_{n{\bf k},\bf G}
 a_{n{\bf k},\bf G'} e^{-i({\bf G}-{\bf G'}+{\bf q}){\bf r}_m}j_1(|{\bf G}-{\bf G'}+{\bf q}|R)\,
 ({\bf G'}+{\bf k}-{\bf q}/2){\cdot} ({\widehat{{\bf G}-{\bf G'}+{\bf q}}})\nonumber\\
+&&\beta^\dagger{\bf s}\beta\;b^*_{n{\bf k},\bf G}
 b_{n{\bf k},\bf G'} e^{-i({\bf G}-{\bf G'}){\bf r}_m}j_1(|{\bf G}-{\bf G'}|R)\,
({\bf G'}+{\bf k}+{\bf q}/2){\cdot} ({\widehat{{\bf G}-{\bf G'}}})\Big],
\label{eq:Qnk}
\end{eqnarray}
\end{widetext}
where $j_1$ is the first spherical Bessel function and $\alpha$, $\beta$ are the up and down spinors respectively. With this expression evaluated it is straight forward to calculate the STT using Eq.~(\ref{eq:flux}) and Eq.~(\ref{eq:A1}).

\section{Materials with spin spiral magnetic order}
\noindent Long ranged magnetic order in form of helical SDWs exists in several types of materials. Maybe the most famous is the heavy rare earth elements.
They have similar valence electron structure, resulting in similar chemical structure, and all the tri-valent elements order in the hcp lattice structure with similar lattice volumes. The main difference along the series lies in the filling and magnetic moment of the localised 4$f$-electron shell. The FS of Er which is typical for the series has a strong nesting feature, i.e. two large parallel sheets of the FS. The formation of a SDW
with a wave vector equal to the nesting vector allows for hybridisation removing these parts of the FS opening up a gap at the Fermi level lowering the total energy of the system. Anisotropies in the system determine whether the type of SDW preferred in the system, is a helical SDW, conical SDW or a longitudinal SDW. 

Helical SDWs also exist for various 3d-transition compounds, however 
among the elemental 3d-transition metals the magnetic structure are less exotic at ambient conditions. It was previously shown that if certain general criteria are met, helical SDWs are formed in the transition metal series. This means that at conditions slightly different from ambient, spin spirals occur naturally indicating that this type of magnetic order is less exotic than one could believe. An example of such a system is $\gamma$-Fe the fcc phase of iron. Experimentally $\gamma$-Fe has been stabilised as precipitates in a Cu-matrix.  The experimental magnetic ground state structure was found to be a SS with $\mathbf{q}=2\pi/a(0.1, 0, 1)$ where $a=$6.76 a.u. Several theoretical studies\cite{Mryasov1992,Bettan,Raquel} have mapped out the ground state magnetic structure of $\gamma$-Fe which was found to be very volume sensitive. In contrast to the rare-earth systems where the SS state is driven by nesting between two parallel sheets of the FS, the SS state of $\gamma$-Fe is not promoted by a single FS nesting vector. Instead there is a net energy gain from many parts of the FS given by the hybridisation by a SS vector.

\begin{figure}
\includegraphics*[width=0.45\textwidth]{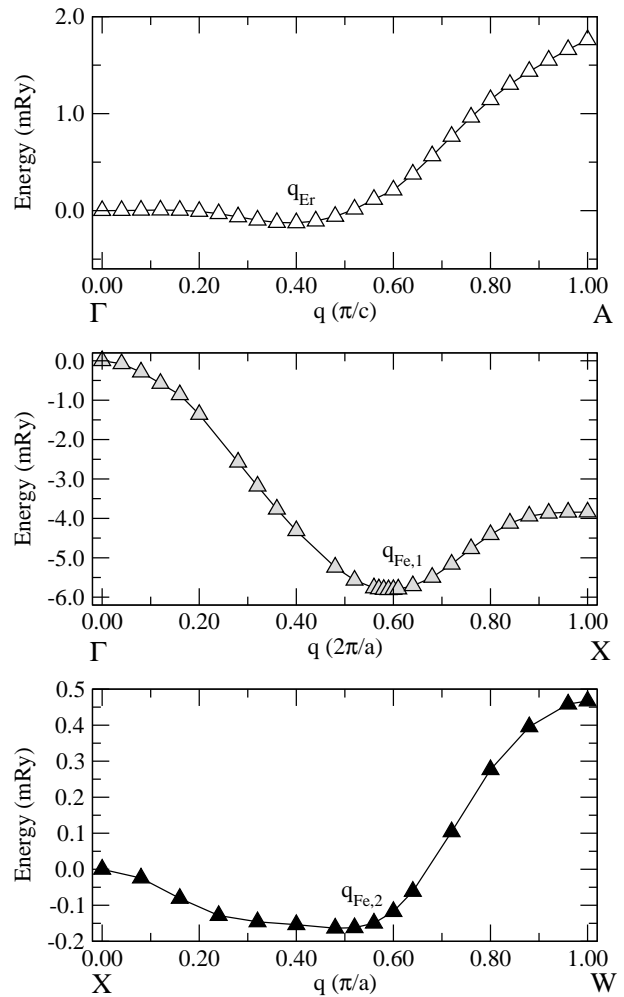}
\caption{The top panel shows the total energy of Er per hcp unit cell for SS wave vectors from $\Gamma$ to $A$. The middle panel shows the total energy per atom of $\gamma$-Fe for SS vectors from $\Gamma$ to $X$ in the larger lattice volume. The bottom panel shows the total energy per atom of $\gamma$-Fe for SS vectors from $X$ to $W$ in the smaller lattice volume.}
\label{fig:es}
\end{figure}

\section{Electronic structure calculation}
\noindent All the materials specific parameters for Er and $\gamma$-Fe used for the calculation of 
the torque current tensor $\textsf{C}$ were obtained from first principles density functional theory. The calculations were made using the full-potential augmented plane wave plus local orbitals (APW+lo) method as described in Ref.~\onlinecite{Bettan2}. For Er, the calculations were performed in the same manner as in Ref.~\onlinecite{Nord,Wessely2006} and for $\gamma$-Fe the calculations were performed in the same way as in Ref.~\onlinecite{Bettan,Raquel}. The local spin density approximation (LSDA) as parametrised by von Barth and Hedin was used without any shape-approximation to the non-collinear magnetisation, i.e. charge and magnetisation densities as well as their conjugates potentials are allowed to vary freely in space both regarding magnitude and direction. For Er, a set of 12$\times$12$\times$8 k-points was used for converging the electron density and for Fe we used a set of 20$\times$20$\times$20 k-points. The SS was treated using the generalised Bloch theorem. The Er calculation was performed with the 4$f$-electrons treated as spin polarised core electrons and we used the experimental Er hcp lattice parameters ($a=6.73$ a.u. and $c=10.56$ a.u.). For Er, we calculate a large number of SS wave vectors, $\mathbf{q}=\pi/c(0,0,q)$ in the first Brillouin zone of the hcp lattice, along the out of plane direction between $\Gamma$ and $A$. The total energy for these calculations are presented in Fig.~\ref{fig:es}. There is an energy minimum for $q=0.40$ which corresponds to the ground state SS structure of Er. For Fe we perform calculations at two lattice volumes. At a lattice constant of $a=6.82$ a.u., corresponding to the lattice constant of Cu, we do a series of calculations for ${\bf q}_1=2\pi/a(0,0,q)$, between $\Gamma$ and $W$ in the first Brillouin zone, and an energy minimum is found for $q=0.59$ (see middle panel in Fig.~\ref{fig:es}). At a reduced lattice constant of $a=6.66$ a.u. we do a series of calculations for ${\bf q}_2=\pi/a(w,0,2)$, between $W$ and $X$, where there is a weak local energy minimum at $w=0.48$. This latter minimum corresponds to a spiral wave vector which is close to the experimental SS of $\gamma$-Fe.
Note that ${\bf q}_2$ is non-parallel to the SS rotation axis which is in the ${\hat {\bf z}}$-direction for all three calculations.  
%For our calculations of the STT we use the Kohn-Sham eigenfunctions obtained at the SS energy minimum.

\section{Results}
\noindent In this section we present the results of the STT calculations for the three SS systems treated in the previous section. 
%For each system was the STT on an atom calculated by summing the contributions from all bands crossing the Fermi level. 
The one electron spin current tensor ${\bf Q}_{n{\bf k}}$ was calculated using Eq.~(\ref{eq:Qnk}) and the Kohn-Sham eigenfunctions obtained from the first principles calculation. The first Brillouin zone was covered by a $41\times 41\times 41$ k-point mesh, in order to accurately evaluate the FS integrals 
%for $\textsf{A}_n$ and $\textsf{B}_n$ 
in Eq.~(\ref{eq:A1}) and Eq.~(\ref{eq:B}). The conductance tensor, $\textsf{B}=\sum_m \textsf{B}_m$, was as expected, found to be diagonal for all three considered systems.

%The spin current density tensor was calculated at the surface of the augmentation spheres of the atoms according to section \ref{sec:apw}. We use a 41$\times$41$\times$41 k-point mesh to cover the first Brillouin zone. 

%\subsection{Erbium}
The torque current tensor $\textsf{C}$ was calculated for Er having a SS with wave vector
${\bf q}=\pi/c(0\; 0\; 0.4)$. 
%We found that the torque current tensor in the ground state spiral structure of Er, ${\bf q}=\pi/c(\,0\, 0\, 0.4\,)$, 
We found that for an Er atom situated at a site with magnetisation direction $(1\, 0\, 0\,)$, the spin current tensor
\[ \textsf{C}
%{\overline {\overline C}}
={\hbar\over e}\left( \begin{array}{ccc}
    0.0 &   0.0 &   0.0\\
    0.0 &   0.0 &   -0.5\\
    0.0 &   0.0 &   0.0\\
\end{array} \right) [\textrm{\AA}^2].\]
From the structure of the
$\textsf{C}$ tensor, it can be seen that for a current in the $(0\, 0\, 1\,)$ direction, a torque is induced in the $(0\, 1\, 0\,)$ direction. Since the torque is perpendicular to the local magnetisation and lies in the SS rotation plane, we conclude that a current along the SS axis causes a rigid rotation of the SS around the SS axis.

%For Er there are four bands that cross the FS. In Table \ref{table:ABPT} we show the individual c$q_2=ontributions to the STT from the different bands. We see that the contribution is largest from the third band. We also show the contributions to $\textsf{A}$ and $\textsf{B}$ from each band.

\begin{table}%[H] add [H] placement to break table across pages
\begin{ruledtabular}
\begin{tabular}{c c c}
a: Er &  &  \\
$n$ & $(\hbar /e)(\textsf{A}_n)_{23}$ [{\AA}eV]& $(\hbar^2 /e^2)(\textsf{B}_n)_{33}$ [eV/{\AA}] \\
\hline
1 & 0.01 & 0.006 \\
2 & 0.01 & 0.01 \\
3 & -0.04 & 0.03 \\
4 & -0.008 & 0.007 \\
\end{tabular}
%\end{ruledtabular}
%\begin{ruledtabular}
\begin{tabular}{c c c}
b: Fe q$_1$ &  & \\
$n$ & $(\hbar /e)(\textsf{A}_n)_{23}$ [{\AA}eV]& $(\hbar^2 /e^2)(\textsf{B}_n)_{33}$ [eV/{\AA}] \\
\hline
1 & -0.002 & 0.002\\
2 & 0.01 & 0.02 \\
3 & -0.1 & 0.07 \\
4 & -0.005 & 0.002 \\
\end{tabular}
%\end{ruledtabular}
%\begin{ruledtabular}
\begin{tabular}{c c c}
c: Fe q$_2$ &  & \\
$n$ & $(\hbar /e)(\textsf{A}_n)_{21}$ [{\AA}eV]& $(\hbar^2 /e^2)(\textsf{B}_n)_{11}$ [eV/{\AA}] \\
\hline
1 & 0.005 & 0.006\\
2 & 0.02 & 0.01\\
3 & 0.02 & 0.02\\
4 & -0.05 & 0.05\\
5 & 0.09 & 0.08\\
\end{tabular}
\end{ruledtabular}
\caption{\label{table:ABPT}The elements of the $\textsf{A}_n$ and $\textsf{B}_n$ tensors that mainly contribute to the $\textsf{C}$ tensor, see Eq.~(\ref{LRJI})}
\end{table}

%\subsection{Iron}
For $\gamma$-Fe with the larger volume in the ${\bf q}_1=2\pi/a(0,0,0.59)$ spiral state, the torque current
tensor, evaluated for an atom with magnetisation direction $(1\, 0\, 0\, )$,% is given by 
\[ \textsf{C}
%{\overline {\overline C}}
={\hbar\over e}\left( \begin{array}{ccc}
    0.0 &   0.0 &   0.0\\
    0.0 &   0.0 &   -1.4\\
    0.0 &   0.0 &   0.0\\
\end{array} \right) [\textrm{\AA}^2].\]
Also in this case we conclude that a current along the SS axis causes a rigid rotation of the SS. In this system, the torque, for the same current density, is
found to be nearly three times larger than for Er. We will return to a discussion on the magnitude of the torque later in this section.
%One reason for the difference in STT 
%is as discussed in connection with Eq.~(6) that the SS wave vector 
%of Er is shorter than for $\gamma$ Fe. Other reasons will be discussed in the 
%end of this section. 

%In the $q_1$ state there are four bands crossing the Fermi level. Contributions to the STT from the four surfaces are given Table \ref{table:ABPT} . The largest contribution comes from the third Fermi surface. 

Finally  we calculated the STT in $\gamma$-Fe with reduced volume, where the SS wave vector ${\bf q}_2=\pi/a(0.48,0,2)$ is non parallel to the spin rotation 
axis ${\hat {\bf z}}$. 
We found that the torque current tensor for an atom with magnetisation in the $(1\, 0\, 0\,)$ direction is given by,
\[ \textsf{C}
%{\overline {\overline C}}
={\hbar\over e}\left( \begin{array}{ccc}
    0.0 &   0.0 &   0.0\\
    0.6 &   0.0 &   0.0\\
    0.0 &   0.0 &   0.0\\
\end{array} \right) [\textrm{\AA}^2].\]
The structure of the torque current tensor for this system differs from the previous two systems. Here, a current in the $(1\, 0\, 0\,)$ direction is required to produce a rotation of the SS. The STT from a current in the $(0\, 0\, 1\,)$ direction vanish since $\gamma$-Fe with ${\bf q}_2=\pi/a(0.48,0,2)$ is antiferromagnetic in the $(0\, 0\, 1\,)$ direction. 

\section{Discussion}
\noindent In general, the STT depends on a balance between the size of the SS wave vector and the spin polarisation of the conduction electrons at the FS. 

Some insight into the origin of the STT on the single electron level can be 
obtained by considering the semi-classical expression for the spin current
\cite{Xiao06}
%[ref Berger JAP 49 2156, Viret EPL 65 427, Stiles PRB 73 54428]
 	\begin{equation}
	    {\bf Q}(z)={\hbar \over 2}{\hat{\bf s}}(z){j\over e}.
	\end{equation}
If the semi-classcal spin current is combined with Eq.~(\ref{eq:torquetensor}) 
%to calculate the torque current tensor 
one obtains the following expression for the torque current tensor.
	\begin{equation}
	  \textsf{C}={\hbar \over 2 e}{\partial {\hat{\bf s}}(z)\over \partial z} 
\Big|_{z_0}dz={ \hbar \over 2e}~\textrm{sin}~\theta_{k_F}\left( \begin{array}{c} -q~\textrm{sin}~(qz_0)~\\
	    q~\textrm{cos}~(qz_0)~\\
	    0\end{array} \right)dz,
	\end{equation}
The above equation gives the same result as the quantum mechanical calculation of $\textsf{C}$ 
in section ~(\ref{sec:1D}), if the denominator of the factor $a$ in 
Eq.~(\ref{eq:afactor}) is equal to one, which is approximately true
if $(q/2k)\textrm{cos}~\theta_{k_F}$ is small.
Semi-classically, the maximum spin transferred per electron, for a given 
value of $q$, to an atomic layer is given by $(\hbar/2)ql$ where $l$
is the inter layer distance. This maximum can be achieved if the single electron 
cone angle $\theta_{k_F}$ is equal to $\pi/2$.
For Er and $\gamma$-Fe with the larger volume is the 
semi-classical maximum of the spin-transfer, with the given value of $q$, equal 
to $0.3\hbar$ and $0.9\hbar$, respectively. 
A more accurate value for the average spin transferred per electron to an atom 
is obtained by multiplying the torque current tensor with $e/A$ where $A$ is the area per atom in the direction of the current. For Er, where the area per atom in the $(0\, 0\, 1\,)$ direction $A=a^2{\sqrt 3}/2$, the spin-transferred per electron is $0.05 \hbar$. For $\gamma$-Fe,  the area per atom in the $(0\, 0\, 1\,)$ direction is $a^2/2$ and the spin-transferred per electron is $0.2 \hbar$,
for the larger volume. Hence, the spin-transfer torques for Er and $\gamma$-Fe (larger volume) obtained from the first principles calculations are only 16\% and 22\%, respectively, of the their semi-classical maximum. This is partially due to the fact 
that the cone angle $\theta_{k_F}$ of the conduction electrons is less than $\pi/2$.

Another reason for the reduced STT is that contributions from different bands partly cancel each 
other. The contributions from the individual bands to the STT are given by the 
 $\textsf{A}_n$ and $\textsf{B}_n$ tensors shown in Table (\ref{table:ABPT}).
%tensor $\textsf{A}_n$, and the conductance $\textsf{B}_n$, are shown in Table (\ref{table:ABPT}). 
For Er there are four bands that cross the FS and the dominant element in the $\textsf{A}_n$ 
tensors is their $(\textsf{A}_n)_{23}$ elements.
As shown in Table (\ref{table:ABPT}a), the STT from band three is four times larger and in the opposite direction 
to the STT from band one and two. For $\gamma$-Fe (larger volume) there are four bands crossing the Fermi level,
where the dominating contribution to the STT comes from the third band, see Table (\ref{table:ABPT}b). 
For $\gamma$-Fe with the smaller volume there are five bands crossing the Fermi level and the dominant element in the $\textsf{A}_n$ tensors is their $(\textsf{A}_n)_{21}$ elements. 
As shown in table (\ref{table:ABPT}c), the largest contribution to the STT is here coming from the fifth FS.
%and there are five bands crossing the Fermi level. The largest contribution to the STT comes from the fourth FS which corresponds to the FS for ${\bf q}_1$ which had the dominant contribution to the torque. 
Finally we would like to note that semi-classically the maximal spintransfer per
conduction electron is $\hbar/2$, which only can be obtained if both 
$ql=\pi/2$ and  $\theta_{k_F}=\pi/2$. An other route to high spin transfer
is to try to reduce the denominator in the expression for the $a$ factor in Eq.~(\ref{eq:afactor}). The $a$ factor becomes large for small Fermi surfaces where $q/k$
is large.

\section{Dynamics}
\label{sec:Dyn}
%In order to understand the magnetisation dynamics of a helical SDW we need to consider the interplay between the STT and magnetic anisotropies. Consider the microscopic equations of motion for atomistic spin dynamics\cite{Skubic2007a} where the time evolution of the spin magnetisation is given by ?the LLG equation (is it called LLG on atomic level)?
\noindent We will for our SS systems represent the uniform magnetisation of an atomic plane
perpendicular to the SS axis by a unit vector ${\hat{\bf m}}$ in the direction of the magnetisation.
The time evolution of ${\hat{\bf m}}$ is phenomenologically described by the LLG equation \cite{Landau,Gilbert}. 
	\begin{equation}
	{d {\hat {\bf m}}\over d t}={\bf \Gamma}-\alpha{\hat{\bf m}}\times{d {\hat {\bf m}}\over d t}.
\label{eq:llg}
	\end{equation}
The first term ${\bf \Gamma}$ is the total torque exerted on the layer and the second term is the Gilbert damping term with damping constant $\alpha$. 
The systems under consideration are all having a helical SDW with an easy-plane or hard-axis anisotropy.
%Now consider a system with a planar SS in a hard-axis anisotropy. 
The hard-axis is directed along the SS axis and helps maintain the planar structure of the SS. For such systems 
is the total torque ${\bf \Gamma}$ a sum of a current induced STT and a torque resulting from the hard-axis anisotropy. Both of these torques are perpendicular to the magnetisation ${\hat{\bf m}}$ and the SS axis which here 
is considered to be in the ${\hat{\bf z}}$ direction.
%both ${\bf z}$ and ${\bf m}$. 
Eq.~(\ref{eq:llg}) can therefore for this system be written as
	\begin{equation}
	{d {\hat {\bf m}}\over d t}=A{\hat {\bf m}}\times{\hat{\bf z}}-\alpha{\hat{\bf m}}\times{d {\hat{\bf m}}\over d t},
\label{eq:llg2}
	\end{equation}
where
	\begin{equation}
	A(j,\theta)={q~j~a(\theta,k_F)\over \textrm{sin}~\theta~}+b\:{\bf m}\cdot{\bf z}.
	\end{equation}
The first term in the above equation is due to the STT and the factor $a(\theta,k_F)$ was
introduced in Eq.~(\ref{eq:torquetensor}). The $b$ in the second term is the strength of the hard-axis 
anisotropy. 
%In order to calculate the time evolution of the magnetisation can from 
Eq.~(\ref{eq:llg2}) is in spherical coordinates, see Fig.~\ref{fig:coord}, given by. 
%The time dependence of the polar and azimuthal angles of the magnetisation are given by
	\begin{equation}
	{d \phi \over d t}(1+\alpha^2)=-A(j,\theta)
\label{eq:polar}
	\end{equation}
and
	\begin{equation}
	{d \theta\over d t}(1+\alpha^2)=-\alpha A(j,\theta)~\textrm{sin}~\theta.
\label{eq:azi}
	\end{equation}
Eq.~(\ref{eq:polar}) describes a rigid rotation of the SS due to the current induced torque where the angular velocity, as a function of the current $j$, is given by $-A(j,\theta)/(1+\alpha^2)$. It can be seen from Eq.~(\ref{eq:azi}) that the effect of the Gilbert damping is to change the cone angle of the SS. Eventually, the spiral will reach a 
steady state with a cone angle $\theta_0$, given by
	\begin{equation}
	{q~j~a(\theta_0,k_F)\over \textrm{sin}~\theta_0~}=-b~\textrm{cos}~\theta_0~,
	\end{equation}
when the current induced torque and the anisotropy torque balance each other.
At this point in time, also the rotational motion of the magnetic moments with respect to the SS axis stops. The polar angle $\phi$ will thus, when a current $j$ is passed along the SS axis, change by
%If initially the spiral is planar then the total change in polar angle before the steady state is reaced is
%in the steady state with  cone angle $\theta_0$ is given by.
	\begin{equation}
	\Delta\phi =-{1\over 1+\alpha^2}\int_{\pi \over 2}^{\theta_0}A(j,\theta)\Big({d \theta\over d t}\Big)^{-1}d\theta
	\end{equation}
or
	\begin{equation}
	\Delta\phi={1\over \alpha}\left[\textrm{ln}\left(\textrm{tan}~{\pi \over 4}~\right)-\textrm{ln}\left(\textrm{tan}~{\theta_0\over 2}~\right)\right],
	\end{equation}
before the spiral reach the steady state with cone angle $\theta_0$. 
%It is interesting to note that the total rotation of the spin spiral due to the applied current 
%is independent of the size of the STT. Instead it 
%is proportional to the inverse of the damping constant. 
The SS will returns to a planar state if
the current is switched off after the steady state is reached. During the reversal, the spiral rotates 
in the opposite direction performing the same number of rotations as required to reach the steady state.

The current induced torque has, for slowly varying magnetisation, been described by introducing an
adiabatic and a non-adiabatic term in the LLG equation \cite{Li,Zhang,Khono,Edwards}. The adiabatic term,
which for a SS whose axis is along ${\hat{\bf z}}$ takes the form $\partial{\hat {\bf m}}/\partial z$,
has in Eq.~(\ref{eq:llg2}) been replased by the STT component of $A(j,\theta)$. A non-adiabatic torque 
term proportional to ${\hat {\bf m}}\times (\partial{\hat {\bf m}}/\partial z)$ could also be introduced in 
Eq.~(\ref{eq:llg2}). The main effect of such a torque would be a modification of Eq.~(\ref{eq:azi}), which
determines the dynamics of the cone angle $\theta$.  A non-adiabatic torque would add a term to the
right hand side of Eq.~(\ref{eq:azi}) which could decrease or even change the sign of $d \theta / d t$.

The effect of in plane anisotropy is discussed in Ref.~\onlinecite{Wessely2006}, where it is suggested that the current needs to overcome a certain critical current before the spiral start to rotate.
%In real systems in-plane anisotropies lead to more complex dynamics. These are discussed in Ref.~\onlinecite{Wessely2006} where it is suggested that these anisotropies lead to a threshold current density required for the SS rotation to start. 

\section{Summary and conclusions}
\noindent We have shown in general, that the a current through a SS induces a rotation 
of the spins, where the angular velocity depends on the magnitude of the 
current. The size of the STT has been analysed in terms of the spin-transfer
per conduction electron and we conclude that the total STT  
depends on, the spin-polarisation of the electrons, the SS wave vector 
and how contributions from different bands coincide. The dynamics of the 
SS has been calculated including the effects of anisotropies 
and damping. For a system with a hard-axis anisotropy along the SS axis, the damping leads to 
a steady state where the rotation eventually stops. 

\acknowledgments
\noindent This work has been supported by grants from the Swedish Research Council (VR)
and the UK Engineering and Physical Sciences Research Council (EPSRC) through the Spin@RT consortium. Calculations have been performed at the Swedish national computer centers HPC2N and NSC. We are also grateful to D. M. Edwards for very helpful conversations.

\bibliography{References}
\end{document}